\documentclass[review]{elsarticle}
\usepackage{lineno,hyperref}
\usepackage{amsmath,amsfonts}
\usepackage{amssymb,latexsym,amsthm, mathrsfs}
\usepackage{graphicx,epstopdf,subcaption}
\usepackage{bm,color}
\newtheorem{theorem}{Theorem}[section]

\newtheorem{exm}{Example}

\modulolinenumbers[5]

\begin{document}

\begin{frontmatter}

\title{Computing stable numerical solutions for multidimensional American option pricing problems: a semi-discretization approach}

\author{Rafael Company
}
\author{Vera Egorova}
\author{Lucas J\'{o}dar}
\author{Fazlollah Soleymani}

\address{Universitat Polit\'{e}cnica de Val\`{e}ncia, camino de Vera s/n, 46022, Valencia, Spain}

{\footnotesize
\begin{abstract}
The matter of the stability for multi-asset American option pricing problems is a present remaining challenge. In this paper a general  transformation of variables allows to remove cross derivative terms reducing the stencil of the proposed numerical scheme and underlying computational cost. Solution of a such problem is constructed by  starting with a semi-discretization approach followed by a full discretization using exponential time differencing and matrix quadrature rules. To the best of our knowledge the stability of the numerical solution is treated in this paper for the first time.
Analysis of the time variation of the numerical solution with respect to previous time level together with the use of   logarithmic norm of matrices are the basis of the stability result. Sufficient stability conditions on step sizes, that also guarantee positivity and boundedness of the solution, are found. Numerical examples for two and three asset problems justify the stability conditions and prove its competitiveness with other relevant methods.
\end{abstract}

}

\end{frontmatter}

\linenumbers

\section{Introduction}

Multi-asset American option pricing problems are frequent and natural in real markets because they satisfy the needs of different investors motivating an increasing interest.
These problems are very challenging due to the complexity, the growing computational cost and also to the difficulty of a serious treatment of the stability.

Numerical methods for valuing multi-asset option pricing of lattice binomial type have been used in \cite{Boyle1989} and \cite{Moon2008}. Numerical methods for pricing multi-asset derivatives by using Monte-Carlo technique are found in \cite{Glasserman2003}. However, in such methods, high computational cost and the fact that the Monte Carlo method only estimates the option value assuming a given starting point and time, motivated the research of alternative approaches.

Fast Fourier transform (FFT) approach, successfully used for one dimensional problems in computational finance by Carr and Madan in \cite{Carr1999}, and has been extended to the multi-asset European case in \cite{Leentvaar2008}, by combining FFT with a sparse grid method. Although the FFT approach does not suffer the so-called curse of dimensionality, it requires that the joint characteristic function of the underlying assets be analytic.

Recently, meshless method had been used to solve multi-asset option pricing problems \cite{Khaliq2004,Shcherbakov2016}, although it has the dimensionality constraint up to three.

Finite difference methods (FDMs) and finite element methods (FEMs) have been widely used to price various derivative securities because they are easy to implement and flexible. Recently multi-asset option pricing problems have been treated using such techniques in \cite{Kovalov2007,Tangman2013,Zhang2015}.

Finite difference approximations have been used for pricing European and American multi-asset options \cite{During2015,Khaliq2004}. Due to opportunity to exercise at any time to maturity, American option pricing problems introduce a free exercise boundary which is more difficult than European options. A common way to treat American options is to write the free boundary problem as a linear complementarity problem (LCP) and then apply method as PSOR to solve it \cite{tavella}.  Here we deal with American options by imposing a small penalty term which allows us to transform the free boundary problem into a nonlinear problem with fixed domain \cite{nielsen,Forsyth2002,Forsyth2007}. In the multi-asset context, due to its versatility, the penalty approach has been combined with several methods such as FDM \cite{Nielsen2D,Khaliq2004}, FEM \cite{Tangman2013} and radial basis functions (RBF) method \cite{Shcherbakov2016,Larsson2016}.

The existence of the cross derivative terms in a PDE makes that the constructed numerical methods, such as FDMs or FEMs, to be more computationally expensive. Furthermore, with respect to numerical analysis viewpoint, such terms may generate oscillations, spurious solutions and other
instabilities \cite{Zvan2003}.
As an illustration, in \cite{During2015} authors proposed a high order compact FDM for solving three-asset  European options obtaining partial stability results. For stochastic volatility models under jump-diffusion processes, in \cite{chiarella2009} authors proposed a special seven-point  approximation of the cross derivative term. Analogous approximation has been suggested in \cite{Khaliq2015} for the multi-asset American option pricing.
The matrix involving the second order partial derivative terms, so called the diffusion matrix, can be diagonalized by means of its orthogonal transformation. This technique could be applied to remove the cross derivative terms as it has been done in \cite{Reisinger2007,Leentvaar2008}.

One interesting approach to solve time-dependent PDEs is the method of lines based on the semi-discretization with respect to spatial variables which results in a system of ordinary differential equations in time with the corresponding matrix of coefficients $A$.
The semi-discretization method has the advantage that it is easy to apply  to multidimensional systems, when one achieves the system of ODEs after the semi-discretization. However, dealing with the analysis of the stability of numerical method a well-known big challenge is to address the stability as the step-size discretizations go to zero, because the size of the matrices tends to infinity.
Recently this technique has been applied in \cite{Tangman2013} to the two-asset  American option model with penalty term. In \cite{MartinVaquero2014} stabilized Runge-Kutta method is proposed for the multi-asset problems based on LCP formulation. Authors use stable methods for the semi-discretized system of ODEs, but
stability of the full discrete numerical scheme for the PDE problem is
not analysed. In \cite{Khaliq2015} a semi-discretized method has been applied for multi-asset problem under regime-switching. In that work the spatial step sizes are fixed, and so the size of the matrix $A$ in order to obtain L-stability.

As the best model may be wasted with careless analysis, the main target of this paper is to  address the stability of finite difference schemes for multi-asset American option pricing problems based on the semi-discretization technique.

We consider an American basket option pricing problem. Let $S_1,\ldots,S_M$ be the asset prices, where $M$ is the number of assets in a portfolio. Let us denote the vector of asset prices $\mathbf{S} = (S_1,\ldots,S_M)^T$ and $P(\mathbf{S}, \tau)$ be the value of American basket option at the moment $\tau$, where $\tau$ is time to maturity $T$, with the payoff function

\begin{equation}\label{payoff}
P(\mathbf{S},0)=\left(E -\sum_{i=1}^{M}\alpha_i S_i \right)^+,
\end{equation}
where $E$ is the strike price and $\alpha_i$ is the positive weight of the corresponding $i$-th asset in the basket.
 Assuming that the asset prices follow a geometric Brownian motion, using Martingale strategies, no-arbitrage principle and It\^{o}'s calculus (see \cite{tavella}), the option price $P(\mathbf{S}, \tau)$ is the solution of the following PDE problem

\begin{equation}\label{multi_BS}
\begin{split}
\frac{\partial P}{\partial \tau}& = \frac{1}{2}\sum_{i=1, j=1}^{M}{\rho_{ij} \sigma_{i} \sigma_{j} S_i S_j\frac{\partial^2 P}{\partial S_i \partial S_j} }+\sum_{i=1}^{M}(r-q_{i})S_i\frac{\partial P}{\partial S_i} -rP+F(P),\\
& S_i>0, \quad i=1,\ldots,M, \quad 0< \tau \leq T,
\end{split}
\end{equation}
where  $\sigma_{i}$ is the volatility of $S_i$, $\rho_{ij}$ is the correlation between $S_i$ and $S_j$, $r$ is the risk-free rate, $q_{i}$ is the constant dividend yield of $i$-th asset.
Let us denote matrix $R \in \mathbb{R}^{M \times M}$ as the correlation matrix with entries $\rho_{ij}$, satisfying $-1\leq\rho_{ij} \leq 1$. The nonlinear penalty term $F(P)$ has several suitable forms \cite{Forsyth2002,Nielsen2D}. Here we chose the following type, see \cite{Khaliq2015,Forsyth2002},
\begin{equation}
F(P) = \lambda\left(P(\mathbf{S},0) - P(\mathbf{S},\tau) \right)^+,
\end{equation}
where $\lambda$ is non-negative.
This penalty term is in accordance with recent rationality parameter approach \cite{Gad2015,our_rationality2016}, that takes into account that the buyer does not exercise when it is not profitable.

Note that at each boundary $S_i=0$ the Black-Scholes equation for $M-1$ assets is established and
\begin{equation}
\lim_{S_i\rightarrow \infty}
P(S_1,\ldots,S_i,\ldots,\tau)=0, \quad 1 \leq i \leq M.
\end{equation}

The rest of the paper is organized as follows. In Section 2 a new transformation of variables based on $LDL^T$ of the correlation matrix proposed in \cite{removing2016} is applied to remove the cross derivative terms. This factorization is based on stable Gaussian elimination and pivoting strategy \cite{Wilkinson,Higham_book,Golub4}, avoiding the computations of eigenvalues and eigenvectors. Once the transformation is applied, the semi-discretization of the transformed PDE results in a system of nonlinear ODEs whose coefficient matrix $A$ and its vector solution are explicitly constructed for general multi-asset case. In Section 3  the resulting semi-discrete system is solved by the accurate Simpson's rule, that allows the integration without imposing invertibility of matrix $A$.
In Section 4 conditional positivity and stability of the solution are shown addressing the growing size of the coefficient matrix with any arbitrarily small step size value. In Sections 5 and 6 the formulations of two-asset and three-asset cases are studied respectively. Moreover, numerical examples are employed in order to put on show the applicability and generality of the proposed method for multi-asset problems alongside comparisons with the existing approaches in the literature. The paper ends with the conclusion section.

\section{Cross derivative term elimination and semi-discretization}

In this section first we apply a dimensionless logarithmic transformation to obtain the multi-asset PDE with constant coefficients. Correspondingly, the initial and boundary conditions will also be changed. Second, we apply $LDL^T$ factorization on the correlation matrix so as to remove the cross derivative terms. Finally, the method of lines is taken into account to semi-discretize the transformed PDE.

Now, we introduce the following dimensionless logarithmic substitution
\begin{equation}\label{transformation}
x_i = \frac{1}{\sigma_i}\ln \frac{S_i}{E}, \; i=1,\ldots,M, \quad V(\mathbf{x}, \tau) = \frac{P(\mathbf{S},\tau)}{E},
\end{equation}
where $\mathbf{x} = [x_1,\ldots,x_M]^T$, that transforms the original PDE (\ref{multi_BS}) into the following form
\begin{equation}\label{multi_BS_log}
\begin{split}
\frac{\partial V}{\partial \tau} = \frac{1}{2}\sum_{i=1, j=1}^{M}{\rho_{ij} \frac{\partial^2 V}{\partial x_i \partial x_j} }+\sum_{i=1}^{M}\delta_i\frac{\partial V}{\partial x_i} -rV +\frac{1}{E}F(EV),\\
x_i \in \mathbb{R}, \quad i=1,...,M, \quad 0<\tau \leq T,
\end{split}
\end{equation}
where $\delta_i = \frac{r-q_{i}-\frac{\sigma_{i}^2}{2}}{\sigma_{i}}$.

By taking advantage of positive semi-definitive property of the correlation matrix $R$, see \cite{Sauer2013,Jewitt2015}, we can apply the stable $LDL^T$ factorization proposed recently by \cite{removing2016}, where $L$ is a unit lower triangular matrix and $D$  is a diagonal matrix with positive diagonal elements $D_{ii}$, such that $R=LDL^T$. Then using the linear transformation
\begin{equation}\label{transformation_y}
\mathbf{y} = [y_1,\ldots,y_M]^T=C\mathbf{x}, \quad U(\mathbf{y},\tau) = V(\mathbf{x},\tau),
\end{equation}
where $C =\left(c_{ij} \right)_{1 \leq i,j \leq M} =  L^{-1}$, equation (\ref{multi_BS_log}) becomes
\begin{equation}\label{multi_BS_cross}
\begin{split}
\frac{\partial U}{\partial \tau} = \frac{1}{2}\sum_{i=1}^{M}{D_{ii}  \frac{\partial^2 U}{\partial y_i^2} }
+\sum_{i=1}^{M}
\left( \sum_{j=1}^{M}
\delta_j
c_{ij}
\right)
\frac{\partial U}{\partial y_i} -rU +\frac{1}{E}F(EU),
\end{split}
\end{equation}
where the cross derivative terms have been removed. Under transformations (\ref{transformation}) and (\ref{transformation_y}) the initial condition (\ref{payoff}) takes the form
\begin{equation}\label{payoff_trans}
U(\mathbf{y},0)=\left(1 - \sum_{i=1}^M \alpha_i e^{\sigma_i x_i} \right)^+,
\end{equation}
where $\mathbf{x}=[x_1,..., x_M]^T = C^{-1}\mathbf{y}$.

Since the numerical solution of the PDE inside a bounded domain will not be crucially affected by the artificial boundary conditions, then some simplified strategies can be taken into consideration, e.g. see Proposition 4.1 in \cite{Jaillet1990}. In this paper we select the artificial boundary conditions at the boundaries of the bounded numerical domain to be equal to the values at $\tau=0$, i.e. the payoff function, for more see \cite{Kovalov2007}.

In order to construct numerical solution, a truncated computational domain has to be considered. Let us chose $y_{i_{min}}$ and $y_{i_{max}}$, $i=1,\ldots,M$ such that boundary conditions are fulfilled. A uniform mesh in each coordinate spatial computational grid of $N_i+1$ nodes with step sizes
$h_i$ takes the following form
\begin{equation}\label{grid}
\xi_i^j = y_{i_{min}}+jh_i, \quad h_i = \frac{y_{i_{max}}-y_{i_{min}}}{N_i},\; 0 \leq j \leq N_i, \; 1 \leq i \leq M.
\end{equation}

An approximate solution at the point $(\xi_1^{j_1},\xi_2^{j_2},\ldots,\xi_M^{j_M},\tau)$ is denoted by $u_{j_1,\ldots,j_M}=u_{j_1,\ldots,j_M}(\tau)$.
Let us denote the set of all mesh points by $\Gamma$, the subset of the mesh points located at the faces of the boundary of the numerical domain by
\begin{equation}
\partial \Gamma = \left\lbrace (\xi_1^{j_1},\xi_2^{j_2},\ldots,\xi_M^{j_M}) \mathrel{}\middle|\mathrel{} \exists m , 1 \leq m \leq M, \; j_m=0 \; \text{or} \; j_m=N_m\right\rbrace,
\end{equation}
the subset of interior nodes by $\Omega = \Gamma \setminus \partial \Gamma $.
Then semi-discretization of the equation (\ref{multi_BS_cross}) is obtained by using the second order central difference approximation for the spatial derivatives,  resulting in the system of nonlinear ordinary differential equations of the form
\begin{equation}\label{2D_MOL}
\begin{split}
\frac{du_{j_1,\ldots,j_M}}{d\tau} & = \frac{1}{2}\sum_{i=1}^{M}{D_{ii} \frac{u_{j_1,\ldots,j_{i}-1,\ldots,j_M} -2 u_{j_1,\ldots,j_{i},\ldots,j_M} +u_{j_1,\ldots,j_{i}+1,\ldots,j_M} }{h_i^2} }\\
&+\sum_{i=1}^{M}
\left( \sum_{j=1}^{M}
\delta_i
c_{ij}
\right)
\frac{u_{j_1,\ldots,j_{i}+1,\ldots,j_M}-u_{j_1,\ldots,j_{i}-1,\ldots,j_M} }{2h_i}\\
& -ru_{j_1,\ldots,j_M} +\frac{1}{E}F(Eu_{j_1,\ldots,j_M} ).
\end{split}
\end{equation}

Note that due to removing transformation the stencil of scheme (\ref{2D_MOL}) is reduced to $2M+1$ mesh points.  In the case of using standard central finite difference approximation of cross derivatives the stencil would be of $2M^2+1$ mesh points   and reduced stencil of \cite{chiarella2009,Khaliq2015} would contain $M^2+M+1$ mesh points.

Let us introduce the following notation for $i=1,\ldots,M$:
\begin{eqnarray}
h_i& = & \beta_i h, \label{hi}\\
d_i &=& \frac{D_{ii}}{\beta_i^2}, \quad d= \sum_{i=1}^{M}d_i, \label{d}\\
c_i  & = & \sum_{j=1}^{M}
\delta_j
c_{ij}, \label{ci}\\
a_0&=&-\frac{1}{h^2}\left( d+rh^2 \right), \label{a0}\\
a_{+i}&=&\frac{1}{2h^2} \left(d_i + \frac{h}{\beta_i}c_i  \right),\\
a_{-i}&=&\frac{1}{2h^2} \left(d_i -\frac{h}{\beta_i}c_i  \right).  \label{ami}
\end{eqnarray}

Let us denote by $\mathbf{u} =\mathbf{u}(\tau) \in \mathbb{R}^{N+1}$  the vector of all values $u_{j_1,\ldots,j_M}$, such that
\begin{equation}
\mathbf{u} = [u_0,\ldots,u_N]^T,
\end{equation}
where $N+1$ means the total number of mesh points, and from (\ref{grid}) one gets
\begin{equation}\label{np1}
N+1 = (N_1+1) (N_2+1)\cdots(N_M+1) = \frac{1}{h^M}\prod_{i=1}^{M} \frac{y_{i_{max}}-y_{i_{min}}+\beta_i h}{\beta_i}.
\end{equation}

Each index $j$, $0 \leq j\leq N$, has a one to one correspondence with the set of indexes $[j_1,\ldots,j_M]$ as follows:
\begin{equation}
[j_1,\ldots,j_M] \equiv  j = j_1+\sum_{m=1}^{M}\left( \prod_{n=1}^{m-1}(N_n+1) \right)j_m.
\end{equation}

Then for index $j$ we denote $\bm{\xi}_j=(\xi_1^{j_1},\xi_2^{j_2},\ldots,\xi_M^{j_M})$. Note that for two indexes $j \equiv [j_1,\ldots,j_M]$ and $i\equiv [i_1,\ldots,i_M]$ the following relations take place. If $i \leq j$, then $i_M \leq j_M$. In the case $i \leq j$ and $i_M = j_M$ one gets $i_{M-1} \leq j_{M-1}$. So, if $i \leq j$ and $i_M = j_M, \ldots ,i_2 = j_M$, then $i_1 \leq j_1$.

System (\ref{2D_MOL}) with the boundary and initial conditions can be presented in the following vector form
\begin{equation}\label{2D_eq_matr}
\begin{cases}
\frac{d\mathbf{u} }{d\tau}(\tau) = A \mathbf{u}(\tau) +\lambda \left(\mathbf{u}(0)-\mathbf{u}(\tau) \right)^+ ,\\
\mathbf{u}(0)= [u_0(0), \ldots, u_N(0)]^T,
\end{cases}
\end{equation}
where
\begin{equation}\label{initial_condition}
u_j(0)= U(\bm{\xi}_j,0) = \left( 1-\sum_{i=1}^{M}\alpha_i e^{\sigma_i x_i(\bm{\xi}_j)}\right)^+,
\end{equation}
where  $x_i(\bm{\xi}_j) = \left( C^{-1} \bm{\xi}_j \right)_i$ is the $i$-th entry of $C^{-1} \bm{\xi}_j$. Matrix $A$ is a sparse banded  $(N+1)\times (N+1) $ matrix  whose  size depends on step size $h$ (see eq. (\ref{np1})), and rows  are entirely with zeros or containing $2M+1$ non-zero entries. In fact,
\begin{equation}
A=(a_{ij})_{0 \leq i,j \leq N},
\end{equation}
\begin{equation}\label{A_matrix}
a_{ij} = \begin{cases}
a_0, & \bm{\xi}_i \in \Omega, \; j =i,\\
a_{\pm 1}, & \bm{\xi}_i \in \Omega, \; j =i\pm 1,\\
a_{\pm m} & \bm{\xi}_i \in \Omega, \; j =i\pm \prod_{n=1}^{m-1}(N_n+1), \; 2 \leq m \leq M, \\
0, & otherwise.
\end{cases}
\end{equation}

Note that as the chosen artificial boundary conditions do not change with $\tau$, then their derivative with respect to $\tau$ are zero which motivates the appearance of zeros in the corresponding rows of $A$. If $\bm{\xi}_i \in \partial \Gamma$, then according to the boundary conditions (\ref{payoff_trans}) the value $u_i(\tau) = u_i(0)$, thus $i$-th equation of the system (\ref{2D_eq_matr}) takes the form
\begin{equation}\label{boundary_condition}
\frac{d u_i(\tau)}{d \tau}=0.
\end{equation}

\section{Full discretization}
In order to solve numerically system (\ref{2D_eq_matr}) we use Exponential Time Differencing (ETD) method \cite{Cox2002}.
Let us introduce temporal discretization with the fixed constant time step $k=\frac{T}{N_{\tau}}$, so $\tau^{n} = nk$, $n=0,\ldots,N_{\tau}$. Then the  exact solution of the system of ODE (\ref{2D_eq_matr}) in some given interval $\tau \in [\tau^n, \tau^{n+1}]$ is  given by  Section 2.1 of \cite{Cox2002}:
\begin{equation}\label{cox_eq}
\mathbf{u}(\tau^{n+1}) = e^{Ak}\mathbf{u}(\tau^{n})+ \lambda\int_{0}^{k} e^{As} \left(\mathbf{u}(0)-\mathbf{u}(\tau^{n+1}-s) \right)^+  ds.
\end{equation}

We propose a first explicit approximation of the integral in (\ref{cox_eq}) by replacing $\mathbf{u}(\tau^{n+1}-s)$ by the known value $\mathbf{u}(\tau^n)$ corresponding to $s=k$. Let us denote $ \mathbf{v}^{n+1}$ by
\begin{equation}
\mathbf{v}^{n+1} = e^{Ak}\mathbf{u}(\tau^n)+ \lambda \left( \int_{0}^{k} e^{As}ds\right)  \left(\mathbf{u}(0)-\mathbf{u}(\tau^n) \right)^+,
\end{equation}
then in accordance with Section 2.1 of \cite{Cox2002}, the local truncation error is

\begin{equation}\label{truncation_error}
\mathbf{u}(\tau^{n+1}) - \mathbf{v}^{n+1} = O(k^2).
\end{equation}

Now instead of solving the integral $\int_{0}^{k} e^{As}ds$ in exact form  involving $A^{-1}$ like \cite{Tangman2013,Khaliq2015}, as matrix $A$ can be singular or ill-conditioned, we use the accurate  Simpson's rule, see \cite{Atkinson},
\begin{equation}
\int_{0}^{k} e^{As}ds = k\varphi(A,k) +O(k^5),
\end{equation}
where
\begin{equation}\label{phi}
\varphi(A,k) =\frac{1}{6} \left(I+4e^{A\frac{k}{2}}+e^{Ak} \right).
\end{equation}

Let $\mathbf{u}^n \approx \mathbf{u}(\tau^n)$ be the numerical solution of the proposed fully discretized explicit scheme
\begin{equation}\label{scheme_1}
\mathbf{u}^{n+1} = e^{Ak}\mathbf{u}^n +  k \lambda \varphi(A,k)\left(\mathbf{u}^0-\mathbf{u}^n\right)^+ , \quad \tau^n = nk, \; n=0, \ldots,N_{\tau}-1.
\end{equation}

 According to (\ref{truncation_error})  and (\ref{phi}) the local truncation error of the full discretized explicit scheme (\ref{scheme_1}) versus the ODE system (\ref{2D_eq_matr}) is of the second order in time.

\section{Positivity and stability}

Next, we pay attention to the stability of the scheme (\ref{scheme_1}) in the classical sense. In fact, we are going to find a step sizes conditions so that the numerical solution of the scheme (\ref{scheme_1}) becomes bounded as the step sizes tend to zero. We also show that the numerical solution is positive. Note that it is not an easy task because the dimension of the matrix $A$ grows as step sizes decrease (see (\ref{grid}) and (\ref{np1})) and the entries of the matrix $A$ also grows (see (\ref{a0})-(\ref{ami})).

For the sake of clarity in the presentation we recall some definitions and results that might be found in \cite{Kaczorek2002}.

A vector $v \in \mathbb{R}^{n}$ (matrix  $A \in \mathbb{R}^{n\times m}$) is called non-negative if its entries $v_{i}$ ($a_{ij}$) are non-negative. The infinite norm is defined by the maximum absolute row sum of the matrix:
\begin{equation}
\left\|A \right\|_{\infty} =\max_{1 \leq i \leq n} \sum_{j=1}^{m} \left| a_{ij}\right|.
\end{equation}
A matrix $A \in \mathbb{R}^{n\times n}$ is called the Metzler matrix if its off-diagonal entries are non-negative:
\begin{equation}
a_{ij} \geq 0, \text{ for } i \neq j, \; i,j =1,2,\ldots,n.
\end{equation}

It is known (see \cite{Kaczorek2002,Arrow1989}) that  if  $A$ is the Metzler matrix, then
\begin{equation}\label{metzler_prop}
e^{At}\geq 0 \text{ for } t\geq 0.
\end{equation}

Further we recall the definition of logarithmic norm, introduced in 1958 independently by Lozinskii \cite{Lozinskii}  and Dahlquist \cite{Dahlquist}.
Let us define an induced (operator) matrix norm   $||\cdot||$ on $\mathbb{C}^{n \times n}$. Then a logarithmic norm of a matrix $A$ is
\begin{equation}
\mu[A] = \lim\limits_{h \rightarrow 0+} \frac{\|I+hA\|-1}{h}.
\end{equation}

The property of the bound of exponential matrix norm by the exponential of the logarithmic norm proposed in \cite{Dahlquist} reads
\begin{equation}\label{prop_log_norm}
\left\| e^{Ak}\right\|  \leq e^{k\mu[A]}.
\end{equation}

The infinity logarithmic norm  can be calculated by using the following formula, see \cite{Feedback}, p. 33,
\begin{eqnarray}
\mu_{\infty}[A]&=&\max_i \left(\Re (a_{ii}) + \sum_{ j \neq i }|a_{ij}| \right), \label{mu3}
\end{eqnarray}
where $\Re(x)$ denotes the real part of complex number $x$.

According to the structure of matrix $A$ of the discretized system (\ref{2D_eq_matr}), described by (\ref{A_matrix}) and (\ref{hi})-(\ref{ami}), and by (\ref{mu3}) the infinity logarithmic norm takes the  form
\begin{equation}\label{mu1A}
\mu_{\infty}[A] = a_0 + \sum_{i=-M, i\neq 0 }^{M}|a_{i}|.
\end{equation}
Coefficients $a_{-i} $ and $a_{+i}$, $i=1,\ldots,M,$ depend on $d_i$ and $c_i$, see (\ref{d}) and (\ref{ci}) respectively. If step size $h$ is chosen as
\begin{equation}\label{h_condition}
h \leq \min_{1 \leq i \leq M}  \frac{d_{i}}{|c_i |},
\end{equation}
then the coefficients $a_{-i} $ and $a_{+i} $ are non-negative. Since matrix $A$ consists of some zero rows, from (\ref{mu1A}) by using the positivity of the coefficients, one gets
\begin{equation}\label{mu1r}
\mu_{\infty}[A] =\max \left\lbrace a_0+ \frac{d}{h^2}, \; 0\right\rbrace =\max\left\lbrace -r, \; 0\right\rbrace =0.
\end{equation}
From (\ref{prop_log_norm}) and (\ref{mu1r}) one gets
\begin{equation}\label{A_bound}
\left\| e^{Ak}\right\|_{\infty}  \leq e^{0} = 1.
\end{equation}
Correspondingly, from (\ref{A_bound}) one gets
\begin{equation}\label{temp3}
\left\| \varphi(A,k)\right\|_{\infty}  \leq \frac{1}{6} \left( 1+4+1\right) \leq 1.
\end{equation}
In fact,
\begin{equation}
\left\| e^{Ak}\right\|_{\infty} = \left\|\varphi(A,k) \right\|_{\infty}=1,
\end{equation}
because $A$ has several zero rows, and their corresponding rows in $e^{Ak}$ have only one entry equal to $1$ while the other entries are zeros.
Now we check that the numerical solution is conditionally non-negative and bounded.

In fact, by (\ref{A_matrix}) diagonal elements of matrix $A$ are zeros and $a_0<0$. By (\ref{h_condition}) all the off-diagonal  elements of $A$ are non-negative, and thus $A$ is a Metzler matrix and by (\ref{metzler_prop}) the exponential $e^{Ak}$ is non-negative. Hence and by (\ref{phi}), $\varphi(A,k)$ is also non-negative. From non-negative $\lambda$ and initial condition $\mathbf{u}^0$, and from (\ref{scheme_1}) the non-negativity of $\mathbf{u}^n$ is established.

Now we prove that $u_i^n \leq 1$, $0 \leq i \leq N$, $0 \leq n \leq N_{\tau}$ by using the induction principle. Note that from (\ref{initial_condition}), $u_i^0 \leq 1$ and from (\ref{scheme_1}) $u_i^{n+1}$ is a function $g_i$ on the arguments $u_0^n,\ldots,u_N^n$, given by
\begin{equation}\label{g_i}
u_{i}^{n+1} = g_i(u_0^n,\ldots,u_N^n)= \left( e^{Ak}\right)_i \mathbf{u}^n+k \lambda \left( \varphi(A,k)\right)_i  \left(\mathbf{u}^0-\mathbf{u}^n \right) ^+.
\end{equation}

Partial derivative of $g_i$ with respect to $u_j^n$ takes the form
\begin{equation}\label{temp2}
\frac{\partial g_i}{\partial u_j^n} =
\begin{cases}
\left(e^{Ak} \right)_{ij}- k \lambda \left(\varphi(A,k) \right)_{ij},  & u_i^0 > u_i^n,\\
\left(e^{Ak} \right)_{ij}, &  u_i^0 \leq u_i^n.
\end{cases} 	
\end{equation}

From non-negativity of $e^{Ak}$ and $\varphi(A,k)$ one gets
\begin{equation}\label{temp04}
\frac{\partial g_i}{\partial u_j^n} \geq \left(e^{Ak} \right)_{ij}- k \lambda \left(\varphi(A,k) \right)_{ij}, \quad 0 \leq i,j \leq N.
\end{equation}
If we denote
\begin{equation}\label{psi}
\Psi(A,k)= e^{Ak} -  k \lambda \varphi(A,k),
\end{equation}
and the vector function $g (u_i^n, \ldots, u_N^n)=[g_1,\ldots, g_N]^T$, then  from (\ref{temp04}) the Jacobian matrix $\frac{\partial g}{\partial u^n}$ satisfies
\begin{equation}
\frac{\partial g}{\partial u^n} \geq \Psi(A,k).
\end{equation}

Note that the non-negativity of $\Psi(A,k)$ guarantees the non-negativity of $\frac{\partial g}{\partial u^n}$ and hence $g_i$ will be increasing in each direction $u_j^n$.

In fact, from (\ref{a0})-(\ref{ami}) and under condition (\ref{h_condition}), $B = A-a_0I$ verifies $B \geq 0$, and taking into account that
\begin{equation}
e^{Ak} = e^{a_0k}e^{Bk},
\end{equation}
$\Psi(A,k)$ can be written as follows
\begin{equation}
\Psi(A,k) = \phi_0(k)I+\sum_{s=1}^{\infty}\phi_s(k)\frac{B^sk^s}{s!},
\end{equation}
where
\begin{eqnarray}
\phi_0(k) &=e^{a_0k}- \frac{k \lambda}{6}\left(1+4e^{a_0 \frac{k}{2}}+e^{a_0k} \right), \label{phi_0}
\\
\phi_s(k)&=e^{a_0k}- \frac{k \lambda}{6}\left(\frac{4}{2^s}e^{a_0 \frac{k}{2}}+e^{a_0k} \right). \label{phi_s}
\end{eqnarray}

Taylor expansion of (\ref{phi_0}) shows that
\begin{equation}\label{phi_0_1}
\phi_0(k) = 1- k(\lambda-a_0)+ k^2\frac{\phi_0''(\xi)}{2} , \quad 0 < \xi<k,
\end{equation}
where
\begin{equation}\label{der_phi}
\phi_0''(\xi) = a_0^2 e^{a_0\xi}+\frac{\lambda}{3}|a_0|e^{a_0 \frac{\xi}{2}}+ \frac{\lambda |a_0|}{6}\left(2-|a_0|\xi \right)\left(e^{a_0\xi}+e^{a_0 \frac{\xi}{2}} \right).
\end{equation}

Note that the sum of the two first terms  of the Taylor expansion of $\phi_0(k) $, $1- k(\lambda-a_0)$ is positive, if
\begin{equation}\label{k_cond1}
k  < \frac{1}{\lambda+|a_0|},
\end{equation}
and by (\ref{a0}) this occurs when
\begin{equation}\label{k_cond}
k  < \frac{h^2}{d+(r+\lambda)h^2}.
\end{equation}

Condition (\ref{k_cond1}) implies  $2-|a_0|\xi>0$ and from (\ref{phi_0_1}) and  (\ref{der_phi}),  $\phi_0''(\xi)$ and $\phi_0(k)$ are positive.
It is easy to check that for $s \geq 1$,$\phi_s(k) \geq \phi_0(k) >0$.
Thus,  Jacobian matrix $\frac{\partial g}{\partial u^n}$ is non-negative, and using induction hypothesis $u_i^n \leq 1$ and (\ref{g_i}) and non-negativity of $u_i^n$, one gets
\begin{equation}
0 \leq u_i^{n+1} = g_i(u_0^n,\ldots,u_N^n) \leq g_i(1,\ldots,1) \leq	  \left\|   \left( e^{Ak}\right) _i\right\|_{\infty}
\leq  \left\|  e^{Ak}\right\|_{\infty} = 1,
\end{equation}
under conditions (\ref{h_condition}) and (\ref{k_cond}).

Summarizing the main result of the paper is established as follows

\begin{theorem}
	With previous notation under conditions (\ref{h_condition}) and (\ref{k_cond}) the numerical solution $\mathbf{u}^n$ of the scheme (\ref{scheme_1}) is non-negative and $\left\| \cdot\right\|_{\infty} $-stable, with $\left\| \mathbf{u}^n \right\|_{\infty} \leq 1 $ for all values of $\lambda \geq 0$ and any time level $0\leq n \leq N_{\tau}$.
\end{theorem}

As a consequence of Theorem 1, and using transformations (\ref{transformation}) and (\ref{transformation_y}), the numerical option price  obtained by the scheme (\ref{scheme_1}) will  take values between zero and strike price $E$. This  fact is in accordance with theory of American basket put option pricing.

\section{Two-asset American basket option pricing}
In this section we consider the case of two underlying assets, i.e., $M=2$.  Then correlation matrix is
\begin{equation}
R=\begin{pmatrix}
1 & \rho\\
\rho & 1
\end{pmatrix} = LDL^T,
\end{equation}
where
\begin{equation}
L= \begin{pmatrix}
1 &0 \\
\rho & 1
\end{pmatrix}, \quad D = diag(1, 1-\rho^2).
\end{equation}

Using  the changes of variables (\ref{transformation}) and (\ref{transformation_y}), equation (\ref{multi_BS}) takes the form
\begin{equation}\label{2D_BS_tr}
\begin{split}
\frac{\partial U}{\partial \tau} & = \frac{1}{2} \frac{\partial^2 U}{\partial y_1^2 } + \frac{1}{2}(1-\rho^2)\frac{\partial^2 U}{\partial y_2^2} + \delta_1 \frac{\partial U}{\partial y_1}\\
& + \left(\delta_2 - \rho \delta_1 \right) \frac{\partial U}{\partial y_2} -rU +\lambda \left(U(\mathbf{y},0)-U(\mathbf{y},\tau) \right)^+ ,
\end{split}
\end{equation}
where  $(y_1,y_2) \in \mathbb{R}^2, \; 0< \tau \leq T,$ and

\begin{equation}\label{transformation2}
y_1 = \frac{1}{\sigma_1} \ln \frac{S_1}{E}, \; y_2 =\frac{1}{\sigma_2} \ln \frac{S_2}{E} -\frac{\rho}{\sigma_1} \ln \frac{S_1}{E}, \; U(y_1,y_2,\tau) = \frac{1}{E} P(S_1,S_2,\tau).
\end{equation}

Initial condition is transformed according to (\ref{transformation2}) in the following form

\begin{equation}
U(y_1,y_2,0) = \left(1-\alpha_1 e^{\sigma_1 y_1} - \alpha_2 e^{\sigma_2(y_2+\rho y_1)} \right)^+.
\end{equation}

Numerical solution is found in bounded domain $[y_{1_{min}},  y_{1_{max}}] \times [y_{2_{min}},  y_{2_{max}}]$. A uniform spatial grid $(\xi_1,\xi_2)$ takes the form (\ref{grid}) with spatial steps $h_i$, denoted by (\ref{hi}). The approximate value of $U(y_1,y_2,\tau)$ at the point $(\xi_1^i, \xi_2^j,\tau)$ is denoted by $u_{i,j}=u_{i,j}(\tau)$. Then,  semi-discretized in space approximation of
equation (\ref{2D_BS_tr}) takes the following five-point stencil form

\begin{equation}
\frac{du_{i,j}}{d \tau} =a_{-2} u_{i,j-1}  + a_{-1}u_{i-1,j}+ a_0 u_{i,j} + a_1 u_{i+1,j} + a_2 u_{i,j+1}  +\lambda  \left( u_{i,j}(0)-u_{i,j}(\tau)\right)^+ ,
\end{equation}
where the coefficients $a_0$ and $a_{\pm i}$ obtained by (\ref{a0})-(\ref{ami}) are

\begin{gather}
a_0 = -\frac{1}{h^2}\left( \frac{1}{\beta_1^2}+\frac{1-\rho^2}{\beta_2^2}+rh^2\right),\quad
a_{\pm 1} = \frac{1}{h^2}\left(\frac{1}{2\beta_1^2} \pm \frac{h\delta_1 }{2\beta_1} \right),\\
a_{\pm 2}   = \frac{1}{h^2}\left(\frac{1-\rho^2}{2\beta_2^2} \pm \frac{h}{2\beta_2}\left( \delta_2 - \rho \delta_1 \right)   \right).
\end{gather}

Next, we present some numerical results. In Example 1 we show that the stability condition (\ref{k_cond}) cannot be removed in the sense that if the condition is broken, the numerical results can be wrong. Furthermore, we compare two algorithms for computing matrix exponentials  in terms of CPU time.

The implementation of the proposed method has been done by using MatLAB R2015a on processor Pentium(R) Dual-Core CPU E5700 3.00 GHz.
The results of the following examples are presented in original variables $(\mathbf{S},\tau)$ obtained by the inverse transformation.

\begin{exm}
We consider American basket put option with no dividends payments pricing with the following parameters
\begin{equation}
\sigma_1 = 0.65, \; \sigma_2 = 0.25, \; r= 0.05, \; \rho = 0.1, \; \alpha_1 = \alpha_2 = 0.5, \; T =1, \; E=9.
\end{equation}
\end{exm}

\begin{figure}[ht]
	\begin{subfigure}[b]{0.5\linewidth}
		\centering
		\includegraphics[width=.9\linewidth]{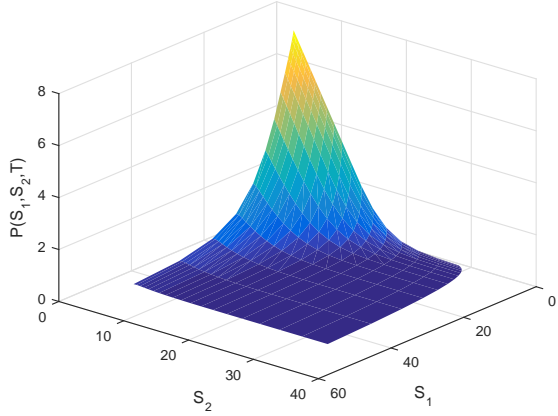}
		\caption{$h=0.5$.}
	\end{subfigure}
	\begin{subfigure}[b]{0.5\linewidth}
		\centering
		\includegraphics[width=.9\linewidth]{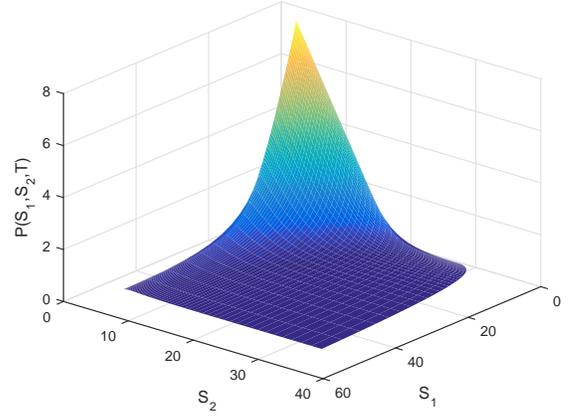}     \caption{$h=0.2$.}
	\end{subfigure}
	\caption{Reliable basket option price of Example 1 at $\tau = T$.}
	\label{fig:example1}
\end{figure}

\begin{figure}[ht]
	\begin{subfigure}[b]{0.5\linewidth}
		\centering
		\includegraphics[width=.9\linewidth]{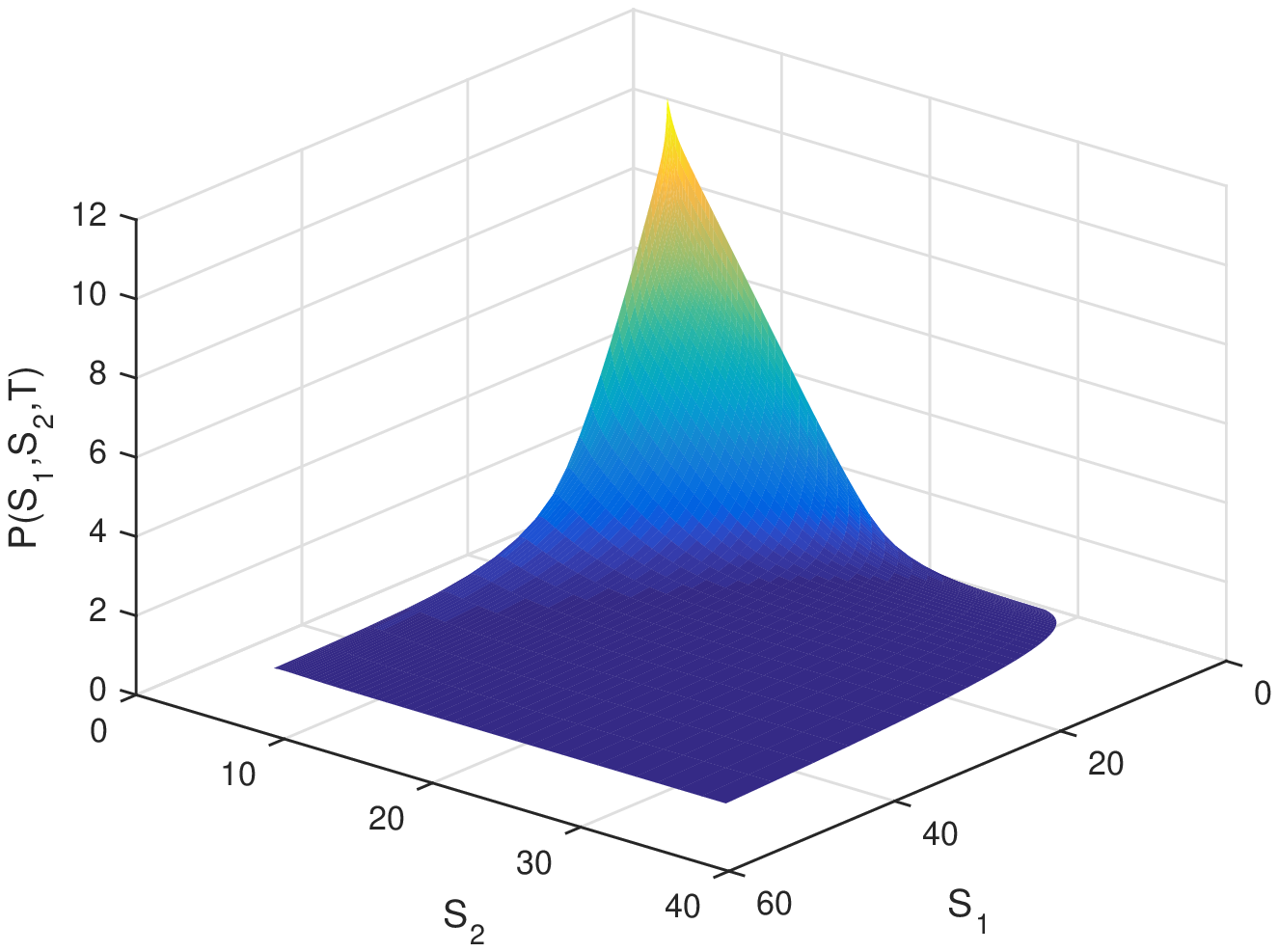}
		\caption{$k=0.05$.}
	\end{subfigure}
	\begin{subfigure}[b]{0.5\linewidth}
		\centering
		\includegraphics[width=.9\linewidth]{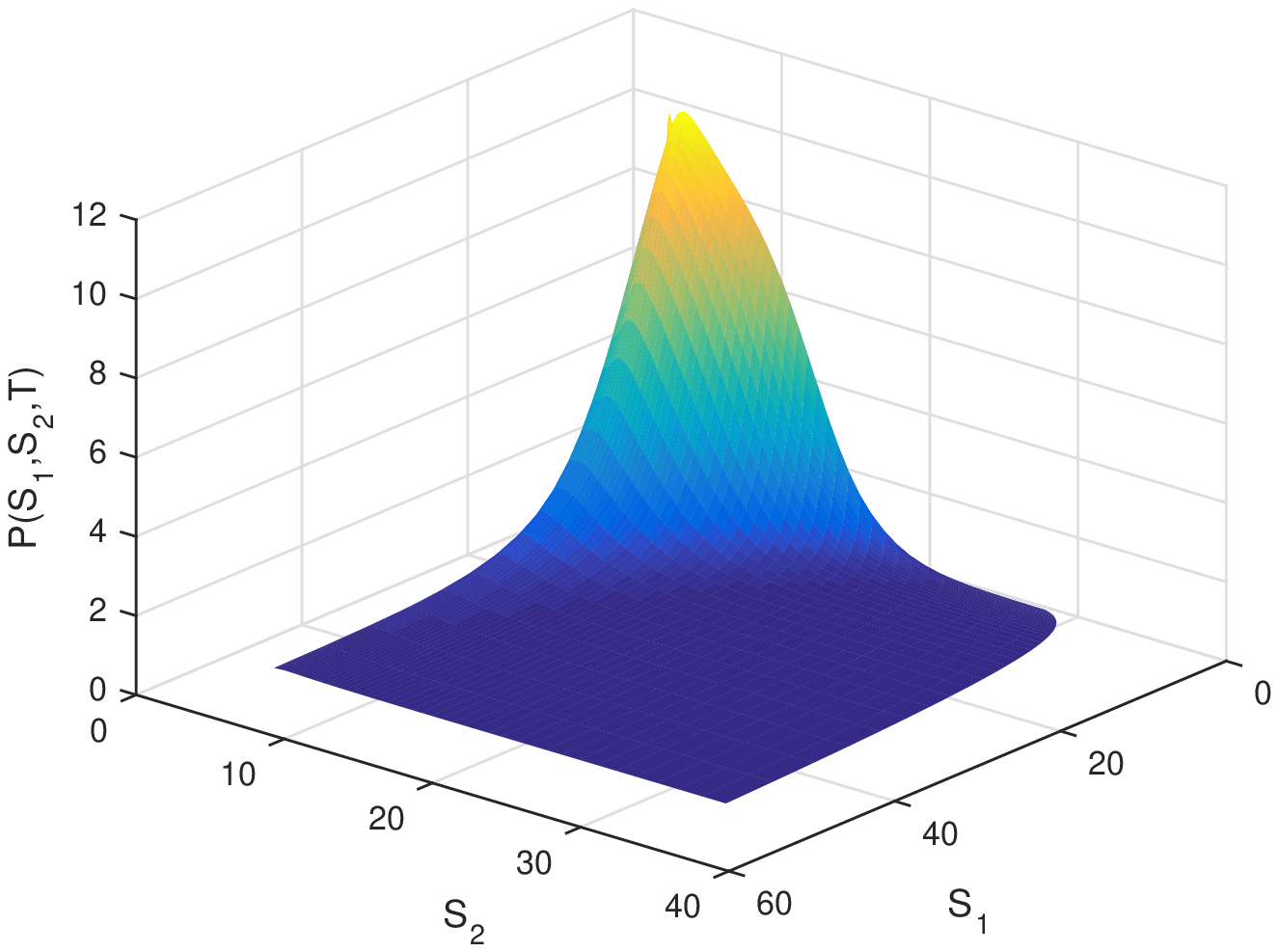}  \caption{$k=0.1$.}
	\end{subfigure}
	\caption{Wrong basket option price of Example 1 at $\tau = T$ with broken stability conditions.}
	\label{fig:example2}
\end{figure}

The penalty parameter is chosen $\lambda=100$, $\beta_1 = \beta_2 = 1$. Transformed computational domain is $[-8,8] \times [-8,8]$. In Fig. \ref{fig:example1} the option price is presented for various $h$ and according to (\ref{k_cond}) fixed $k=8\cdot 10^{-3}$. If time step is chosen larger, for example, $k=0.05$ or $k=0.1$ (see Fig. \ref{fig:example2}), the solution exceeds the strike value $E$, which is unsuitable.

In the computation of numerical solution, the matrix exponential function is very time consuming. In MatLAB library there is a function for a such computation based on algorithm of high performance computing of the matrix exponential proposed in \cite{Higham_algorithm}. Recently, an alternative algorithm in high performance computing for matrix exponential is  proposed in \cite{emilio}. Both approaches for the fixed time step $k=5 \cdot 10^{-4}$ and various spatial steps $h$ are compared in Table \ref{table:CPU1}.

\begin{table}[h]
	\begin{center}
		\begin{tabular}{|c|cc|cc|}
			\hline
			$h$ & \multicolumn{2}{c|}{ \cite{Higham_algorithm}} &  \multicolumn{2}{c|}{ \cite{emilio}}  \\
			\hline
			& Full method & Matrix exp.& Full method & Matrix exp.\\
			\hline
			0.5 & 0.541 & 0.204 & 0.548 & 0.218 \\
			0.2 & 34.617 & 28.608 & 31.150 & 24.622 \\
			0.15 & 286.740 & 256.146 & 190.405 & 167.399\\
			\hline
		\end{tabular}
		\caption{CPU-time (in sec.) of the proposed method itself and matrix exponential by using algorithms of \cite{Higham_algorithm} and \cite{emilio} for Example 1.}
		\label{table:CPU1}
	\end{center}
\end{table}

In Example 2 results are compared with the penalty method without cross derivative term elimination proposed in \cite{Khaliq2015} and the tree method of \cite{Borovkova2012}. Dependence of the solution on the parameter $\lambda$ is also studied.

\begin{exm}
	The American basket put option of two assets is considered with the following parameters \cite{Borovkova2012}
	\begin{equation}\label{exm_borovkova}
	\sigma_1 = 0.3, \; \sigma_2 = 0.2, \; r= 0.05, \; \rho = 0.6, \; \alpha_1 = 0.7, \; \alpha_2 = 0.3, \; T =1, \; E=50.
	\end{equation}
\end{exm}

As a reference value at the point $\mathbf{S} = (50, 50)$ the result of the Binomial Tree method of \cite{Borovkova2012} is used. The results of the proposed method with various spatial step sizes $h$ and fixed $k=5 \cdot 10^{-3}$, in the computational spatial domain $[-8, 8] \times [-8,8]$ are compared with the method of \cite{Khaliq2015} (KM), when cross derivative terms have not been removed, in Table \ref{table:Comparison1}.

\begin{table}[h]
\begin{center}
\begin{tabular}{|c|c|cc|cc|}\hline
$h$	& Number of nodes & \multicolumn{2}{c|}{ Proposed method ($P_h$)} & \multicolumn{2}{c|}{KM}\\
\hline
& & Value & Ratio & Value & Ratio \\
0.8 & $21 \times 21$ & 3.7075 &  &3.8840 & \\
0.4 & $41 \times 41$ & 3.9537 & 12.5047  &  3.9543 & 4.3735\\
0.2 & $81 \times 81$& 3.9730 & 10.1905 &3.9552 &  1.0467\\
0.1 &$161 \times 161$ & 3.9747  & 5.2500  &3.9546 & 0.9722 \\\hline
\multicolumn{2}{|c|}{Tree method ($P$)}& \multicolumn{4}{c|}{3.9751}\\\hline
\end{tabular}
\caption{Comparison of option price for Example 2.}
\label{table:Comparison1}
\end{center}
\end{table}

\begin{figure}[ht]
\centering
\includegraphics[width=.7\linewidth]{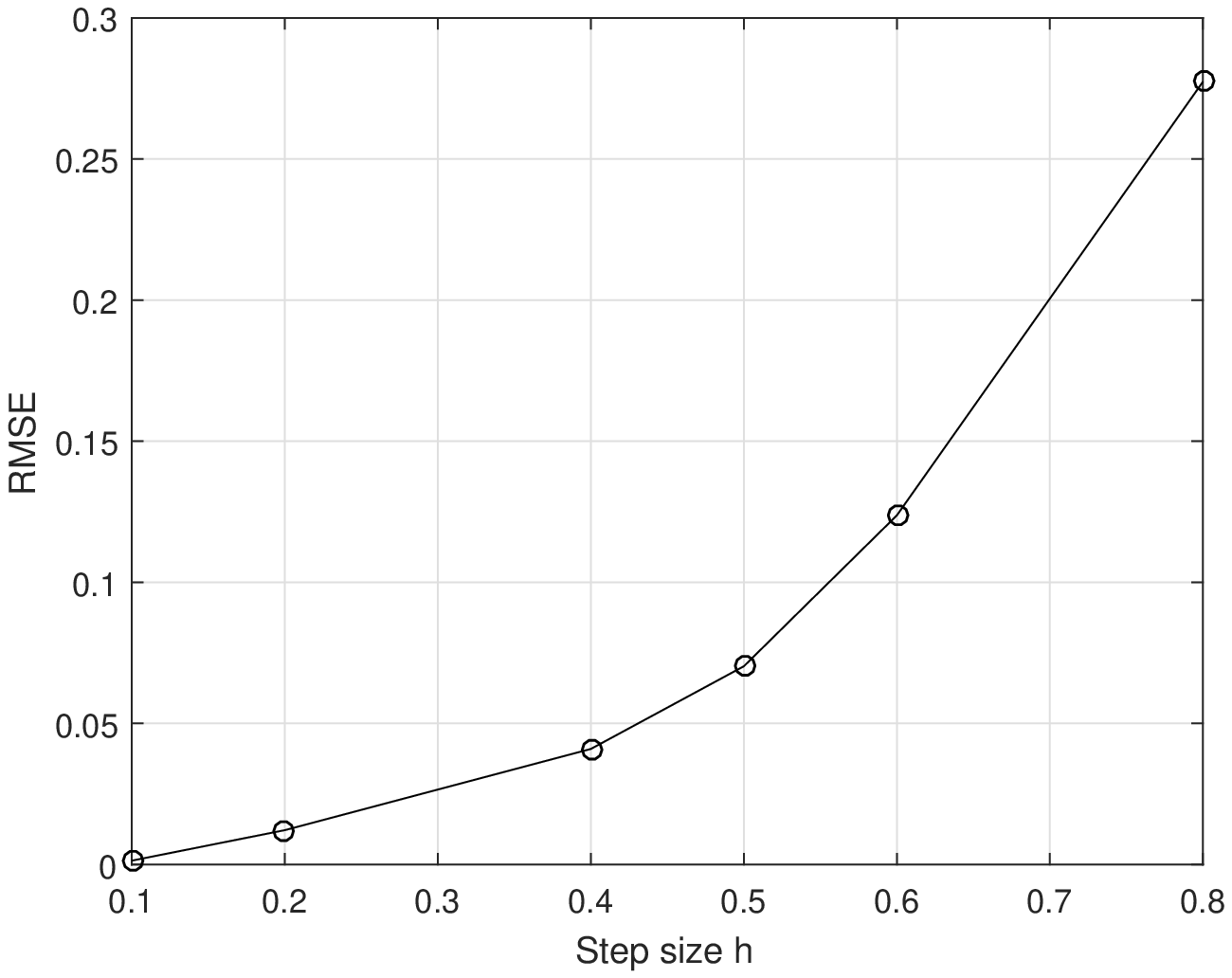}
\caption{Absolute error $\epsilon_h$ of the proposed method  in Example 2 for various $h$ .}
\label{fig:RMSE}
\end{figure}

The convergence ratio that is the factor by which the error  decreases at each grid
refinement is also presented in Table \ref{table:Comparison1}, where the absolute error is computed as follows
\begin{equation}\label{RMSE}
\epsilon_h = \left|  P_h - P\right| ,
\end{equation}
where $P_h$ is the computed value of the option, $P$ is the reference value obtained by the tree method in \cite{Borovkova2012}. The error $\epsilon$ is plotted for various step sizes $h$ in Figure \ref{fig:RMSE}.

The choice of time step $k$  depends on the value of the parameter $\lambda$. In Table \ref{table:lambdas} values of the basket option with parameters (\ref{exm_borovkova}) at $\mathbf{S}=(50, 50)$ applying fixed spatial step size $h=0.2$ are presented depending on $\lambda$.

\begin{table}[h]
\begin{center}\begin{tabular}{|c|c|}\hline
$\lambda$ & $P_h$  \\\hline
0 & 3.6583  \\
1 & 3.7869    \\
10 & 3.9288    \\
100 &3.9730    \\
1000 &3.9732   \\
10000 &3.9733   \\\hline
Tree method ($P$) & 3.9751\\\hline
\end{tabular}
\caption{Option price for the parameters (\ref{exm_borovkova}).}
\label{table:lambdas}
\end{center}
\end{table}

The numerical simulations of Example 2 show that the accuracy remains almost fixed for values of $\lambda>100$. It is advisable to chose $\lambda$ about $100$ to save the computational time.

\vspace{0.5cm}
The proposed method can be applied not only for put options, but also for call options. The payoff function (\ref{payoff}) in this case takes the following form
\begin{equation}\label{payoff_call}
P(\mathbf{S},0)=\left( \sum_{i=1}^{M}\alpha_i S_i  - E\right)^+.
\end{equation}

Example 3 provides numerical solution for American basket call option and its comparison with high order finite element method of \cite{Tangman2013}.

\begin{exm}
The American basket call option of two assets is considered with the following parameters \cite{Tangman2013}
\begin{equation}\label{exm_tangman}
\sigma_1 = 0.12, \; \sigma_2 = 0.14, \; r= 0.03, \; \rho = 0.3, \; q_1 = 0.01, \; q_2 = 0.01, \; T =0.5, \; E=100.
\end{equation}
\end{exm}

\begin{figure}[ht]
	\begin{subfigure}[b]{0.5\linewidth}
		\centering
		\includegraphics[width=.9\linewidth]{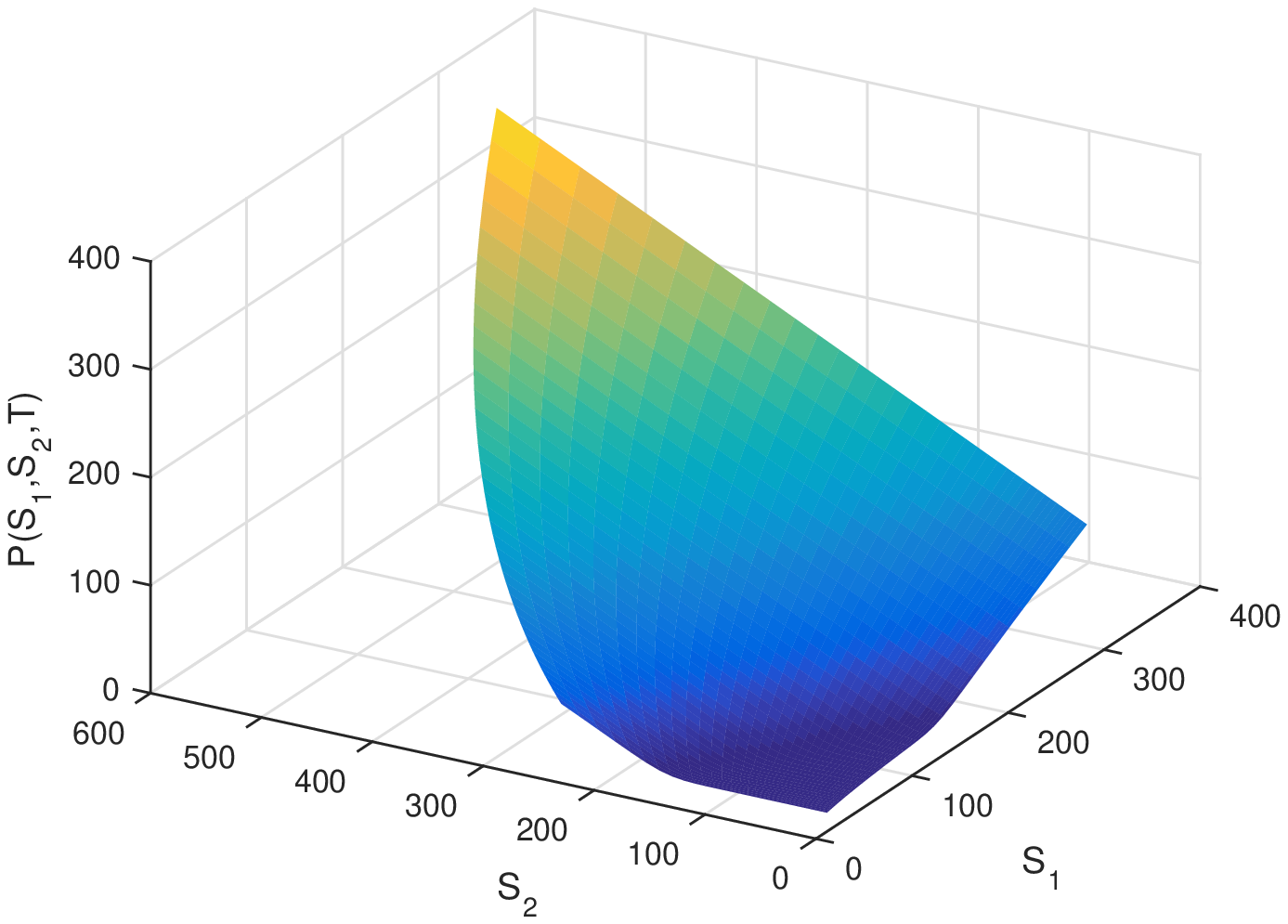}
		\caption{$\tau=T$.}
	\end{subfigure}
	\begin{subfigure}[b]{0.5\linewidth}
		\centering
		\includegraphics[width=.9\linewidth]{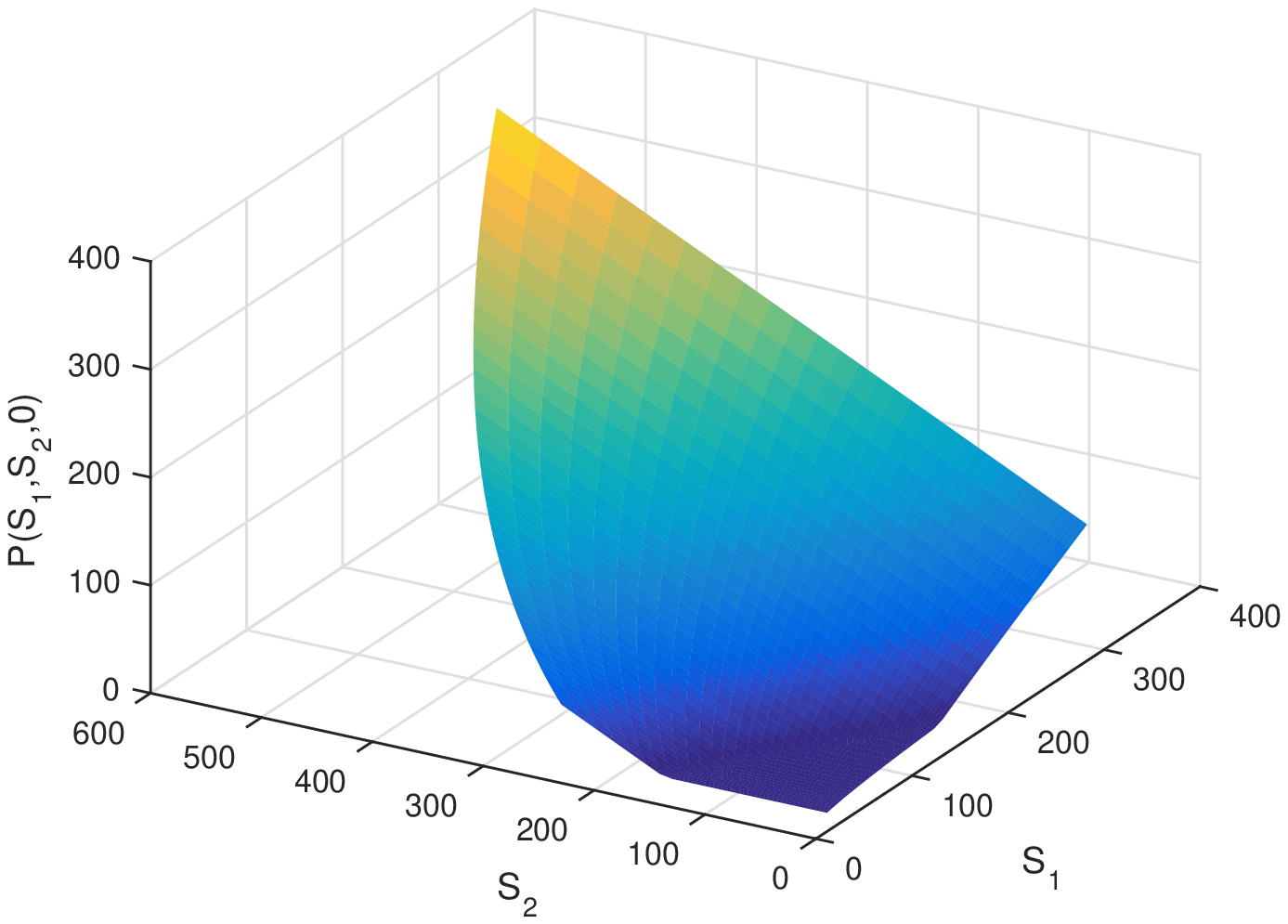}     \caption{$\tau=0$.}
	\end{subfigure}
	\caption{Basket call option price with parameters (\ref{exm_tangman}).}
	\label{fig:call2D}
\end{figure}

In Table  \ref{table:call2D} we include the results at $\mathbf{S} = (100, 100)$ for   $\lambda=100$, various spatial step sizes $h$ and corresponding $k$ under condition (\ref{k_cond}). The numerical solution by high-order computational method of \cite{Tangman2013} is denoted by HOC. The numerical solution at $\tau=T$ and the payoff for American basket call options  are presented in Fig. \ref{fig:call2D}.

\begin{table}[h]
	\begin{center}
		\begin{tabular}{|c|c|c|}
			\hline
			Nodes & Proposed method & HOC   \\
			\hline
			12 $\times$ 12 & 3.18982 &2.86247  \\
			24 $\times$ 24 & 3.35338 & 3.27894\\
			48 $\times$ 48 & 3.41344 & 3.35094  \\
			\hline
		\end{tabular}
		\caption{American basket call option price comparison for Example 3.}
		\label{table:call2D}
	\end{center}
\end{table}

Next we apply the proposed method to the American basket option on three assets. However, numerical example is provided for European option in order to compare it with sparse grid solution technique of \cite{Leentvaartesis}.

\section{Three-asset American Basket option}
The considered approach of cross derivative elimination  can be applied to any multi-dimensional Black-Scholes equation.  In the case of three underlying assets, transformation matrix $C$ with preliminary logarithmic transformation (\ref{transformation}) results in the following new variables

\begin{equation}\label{transformation_3D}
\begin{split}
y_1  & =  \frac{1}{\sigma_1}\ln \frac{S_1}{E},\\
y_2 &  = \frac{1}{\sigma_2} \ln \frac{S_2}{E} -\frac{\rho_{12}}{\sigma_1} \ln \frac{S_1}{E}, \\
y_3 & =  \frac{1}{\sigma_3}\ln \frac{S_3}{E} +
\frac{\beta}{\sigma_2}\ln \frac{S_2}{E}-\left( \beta \rho_{12}+\rho_{13}\right)\frac{1}{\sigma_1}\ln \frac{S_1}{E},
\\ U&(y_1,  y_2,y_3,\tau)  = \frac{1}{E} P(S_1,S_2,S_3,\tau),
\end{split}
\end{equation}
where $\beta = \frac{\rho_{12} \rho_{13}-\rho_{23}}{1-\rho_{12}^2}$. Applying substitution (\ref{transformation_3D}) to equation (\ref{multi_BS}), one gets

\begin{equation}\label{3D_equation}
\begin{split}
\frac{\partial U}{\partial \tau} &= \frac{1}{2} \frac{\partial^2 U}{\partial y_1^2}
+\frac{1-\rho_{12}^2}{2}\frac{\partial^2 U}{\partial y_2^2}
+\frac{\det R}{2 (1-\rho_{12}^2)}\frac{\partial^2 U}{\partial y_3^2}\\
&+ \delta_1 \frac{\partial U}{\partial y_1}  + \left[\delta_2 -\rho_{12}\delta_1\right]\frac{\partial U}{\partial y_2}  +\left[\delta_3
+\beta\delta_2
-\left( \beta \rho_{12}+\rho_{13}\right) \delta_1\right]\frac{\partial U}{\partial y_3}-rU.
\end{split}
\end{equation}

Payoff function for basket call option (\ref{payoff_call}) in new variables takes the following form
\begin{equation}
U(\mathbf{y},0) = \left(\alpha_1 e^{\sigma_1 y_1}+\alpha_2 e^{\sigma_2(y_2+\rho_{12}y_1)}+\alpha_3 e^{\sigma_3\left(y_3-\beta y_2 + \rho_{13}y_1 \right) } -1\right)^+ .
\end{equation}

\begin{figure}[ht]
	\centering
	\includegraphics[scale=0.6]
	{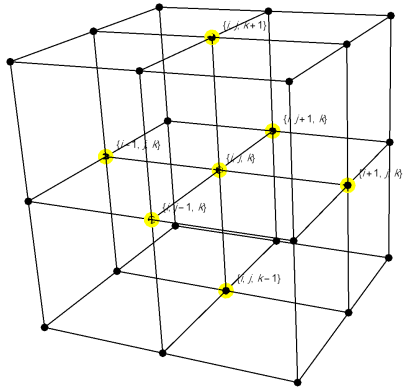}
	\caption{The seven-point stencil for the 3D case.}
	\label{fig3dpic}
\end{figure}

Then semi-discretization of  (\ref{3D_equation}) takes the following seven-point stencil form (see Fig. \ref{fig3dpic}),

\begin{equation}
\begin{split}
\frac{du_{i,j,l}}{d \tau} &=a_{-3}u_{i,j,l-1}+a_{-2} u_{i,j-1,l} + a_{-1}u_{i-1,j,l}+ a_0 u_{i,j,l}\\
& + a_1 u_{i+1,j,l} + a_2 u_{i,j+1,l} + a_{-3}u_{i,j,l-1} +\lambda \left(u_{i,j,l}(0)-u_{i,j,l}(\tau) \right)^+ ,
\end{split}
\end{equation}
where the coefficients $a_0$ and $a_{\pm m}$, $m=1,2,3$, are obtained by (\ref{a0})-(\ref{ami}).

\begin{exm}
As a numerical example we consider the European basket call option with no dividends and the following parameters (see \cite{Leentvaartesis}, p. 76)
\begin{equation}\label{3D_parameters}
\sigma_1 =0.3, \; \sigma_2 = 0.35, \; \sigma_4 = 0.4,\; r = 0.04, \; \rho_{ij} = 0.5, \alpha_i=\frac{1}{3}, \; T=1, \; E=100.
\end{equation}
\end{exm}

The spot price is chosen to be $S_1=S_2=S_3=E$. The reference value $P_{ref}=13.245$ is computed by using an accurate Fast Fourier Transform technique (see \cite{Leentvaartesis}, chapter 4). Since the considered option is of European style, penalty term is not necessary and $\lambda$ is chosen to be zero. The numerical results of the proposed method $P_h$ are presented in the following table and compared with the sparse grid solution technique $P_l$ on an equidistant grid of \cite{Leentvaar2008JCAM} and the method of \cite{Khaliq2015} denoted by KM with rationality approach \cite{Gad2015}.

\begin{table}[h]
	\begin{center}
		\begin{tabular}{|c|c|c|c|}
			\hline
			$n$ & \multicolumn{1}{c|}{$P_h$}  & \multicolumn{1}{c|}{$P_{l}$}& \multicolumn{1}{c|}{KM (with rationality)}\\
			\hline
			8&    11.4957  & {12.8618} & 12.394\\
			16 & 13.3457 & {13.1501}   & 13.055 \\
			32 & 13.3272 & {13.2214} & 13.235 \\
			64 & 13.2470 & {13.2390} &13.241\\
			\hline
			Reference value ($P$) & \multicolumn{3}{c|}{{13.2449}}\\
			\hline
		\end{tabular}
		\caption{Option price on an equidistant grid of $n\times n \times n$ nodes.}
		\label{table:3D}
	\end{center}
\end{table}

\section{Conclusions}
To the best of our knowledge in this paper the stability of numerical solution of multi-asset American option pricing problems is  treated by first time. Change of variables based on $LDL^T$ factorization of the correlation matrix results in the  elimination of cross derivative terms  that allowing the reduction of the stencil of difference scheme and saving the computational cost. After spatial semi-discretization, the problem is fully discretized and using logarithmic norm  of matrices, exponential time differencing ideas and properties of matrix exponential, sufficient condition on the step sizes are given so that the numerical solution of the difference scheme remains norm bounded as the step sizes tend to zero.  Moreover, these conditions are sufficient for the positivity of the solution, that is important dealing with prices of derivatives.

This paper clarifies at once the confusion developed by some authors that dealing with the stability of the solution of a numerical scheme, talking about the stability of the solution of the system of ODEs achieved after semi-discretization, or frozen the size of matrices fixing some step sizes, or argue that stability of schemes for ordinary differential equations instead of the fully discretized scheme of the multi-asset PDE problem.  Results are illustrated with numerical examples for  two-asset and three-asset basket put and call options. Comparison with other relevant methods shows the competitiveness of the proposed method.

\section*{Acknowledgements}
This work has been partially supported by the European Union in the FP7-PEOPLE-2012-ITN program under Grant Agreement Number 304617 (FP7 Marie Curie Action, Project Multi-ITN STRIKE-Novel Methods in Computational Finance) and the Ministerio de Econom\'{\i}a y Competitividad Spanish grant MTM2013-41765-P.

{\footnotesize
\bibliography{references}
}

\end{document}